\documentclass[useAMS,usegraphicx]{mn2e}
\usepackage{natbib}
\usepackage{amsmath,amssymb}
\usepackage{url}
\usepackage{multirow}
\bibpunct[; ]{(}{)}{;}{a}{}{,}

\newcommand{\hatlas}{\mbox{H-ATLAS}}
\newcommand{\sigmapos}{\ensuremath{\sigma_\text{pos}}}
\newcommand{\mum}{\ensuremath{\mu}\text{m}}
\newcommand{\Herschel}{\emph{Herschel}}
\newcommand{\hour}{\ensuremath{^\text{h}}}
\newcommand{\eqnref}[1]{equation~(\ref{#1})}

\defcitealias{Smith2011a}{S11}

\title[Colour and positional offsets of SMGs]{Colour matters: the effects of lensing on the positional offsets between optical and submillimetre galaxies in \Herschel\thanks{\Herschel\ is an ESA space observatory with science instruments provided by European-led Principal Investigator consortia and with important participation from NASA.}-ATLAS}
\author[N. Bourne et al.]
  {N.~Bourne,$^{1,2}$\thanks{Email: nbourne22@gmail.com}
   S.\,J.~Maddox,$^{2,3}$
   L.~Dunne,$^{2,3}$
   S.~Dye,$^1$
   S.~Eales,$^4$   
   C.~Hoyos,$^{1,5}$ \newauthor 
   J.~Gonz{\'a}lez-Nuevo,$^6$
   D.\,J.\,B.~Smith,$^7$
   E.~Valiante,$^4$
   G.~de~Zotti,$^{8,9}$   
   R.\,J.~Ivison,$^{2,10}$\newauthor 
   K.~Rowlands$^{11}$
   \\
   $^1$School of Physics and Astronomy, University of Nottingham, University Park, Nottingham, NG7 2RD, UK\\
   $^2$Institute for Astronomy, University of Edinburgh, Royal Observatory, Edinburgh EH9 3HJ, UK\\
   $^3$Department of Physics \& Astronomy, University of Canterbury, Christchurch, New Zealand\\ 
   $^4$School of Physics and Astronomy, Cardiff University, The Parade, Cardiff, CF24 3AA, UK\\   
   $^5$Instituto de Astronomia, Geof{\'i}sica e Ci{\^e}ncias Atmosf{\'e}ricas, Universidade de S{\~a}o Paulo, Rua do Mat{\~a}o, 1226, Cidade Universit{\'a}ria \\S{\~a}o Paulo, Brazil 05508-090\\
   $^6$Inst. de Fisica de Cantabria (CSIC-UC), Avda. los Castros s/n, 39005 Santander, Spain\\
   $^7$Centre for Astrophysics, Science \& Technology Research Institute,	University of Hertfordshire, Hatfield, Herts, AL10 9AB, UK \\
   $^{8}$INAF-Osservatorio Astronomico di Padova, Vicolo Osservatorio 5, I-35122 Padova, Italy\\
   ${^9}$SISSA, Via Bonomea 265, I-34136 Trieste, Italy\\
   $^{10}$ESO, Karl Schwarzschild Strasse 2, D-85748 Garching, Germany\\
   $^{11}$(SUPA) School of Physics \& Astronomy, University of St Andrews, North Haugh, St Andrews, KY16 9SS, UK\\
    }

\pagerange{\pageref{firstpage}--\pageref{lastpage}} 
\pubyear{2014}

\begin{document}

\label{firstpage}

\maketitle

\begin{abstract}
We report an unexpected variation in the positional offset distributions between \textit{Herschel}-ATLAS sub-millimetre (submm) sources and their optical associations, depending on both 250-\mum\ signal-to-noise ratio and 250/350-\mum\ colour. 
We show that redder and brighter submm sources have optical associations with a broader distribution of positional offsets than would be expected if these offsets were due to random positional errors in the source extraction.
The observation can be explained by two possible effects: either red submm sources trace a more clustered population than blue ones, and their positional errors are increased by confusion; or red submm sources are generally at high redshifts and are frequently associated with low-redshift lensing structures which are identified as false counterparts.
We perform various analyses of the data, including the multiplicity of optical associations, the redshift and magnitude distributions in \hatlas\ in comparison to HerMES, and simulations of weak lensing, and we conclude that the effects are most likely to be explained by widespread weak lensing of \Herschel-SPIRE sources by foreground structures.
This has important consequences for counterpart identification and derived redshift distributions and luminosity functions of submm surveys. 
\end{abstract}

\begin{keywords}
galaxies: statistics --
submillimetre: galaxies --
gravitational lensing: weak
\end{keywords}

\section{Introduction}
The field of sub-millimetre (submm) astronomy has seen huge advances over recent years, but one of the most difficult and persistent issues is unambiguous identification of extragalactic sources. 
Thanks to the negative k--correction, sensitive instruments such as JCMT/SCUBA-2 \citep{Holland2013} and \textit{Herschel}/SPIRE \citep{Pilbratt2010,Griffin2010} are able to detect galaxies over a broad range of redshifts, but this is a double-edged sword when the resulting confusion in the low-resolution images leads to ambiguity in matching with sources identified at other wavelengths \citep[etc]{Hughes1998,Barger1999,Smail2000,Ivison2002,Chapman2003b}.
The confusion problem is exacerbated by galaxy clustering, which increases the chances of galaxies at the same redshift being blended together in the submm. It can also be compounded by gravitational lensing, which magnifies background sources when a foreground galaxy, group or cluster is close to the line of sight, hence increasing the surface density of background sources above a given flux limit in that line of sight \citep{Blain1996,Smail1997,Aretxaga2011,Noble2011}. When matching submm sources to optical/near-infrared surveys it is likely that the background source will be detected in the submm, while only the foreground lens may appear in the reference survey, so that a false counterpart may be assigned \citep{Negrello2007,Negrello2010,Gonzalez-Nuevo2012,Wardlow2012}. 
Hence lensing has the potential to affect the statistics of (apparent) multi-wavelength counterparts to submm sources \citep{Chapin2011}. Blending and clustering can also affect these statistics if they change the positional errors of sources extracted from the submm images \citep{Hogg2001,Negrello2005}. It is essential to quantify these effects in order to fully understand the source counts, redshifts and luminosity functions of submm galaxies (SMGs).
In the current paper we focus on the observed positional offsets between SPIRE 250-\mum\ sources and optical associations which do not follow the expected behaviour. The measurements are described in Section~\ref{sec:data}, and in Section~\ref{sec:results} we present new evidence for a dependence on submm colour and signal-to-noise ratio (SNR). We explore the potential interpretations for these observations in terms of clustering and lensing, and the implications of each, in Section~\ref{sec:discussion}.

\section{\hatlas/SDSS positional offsets}
\label{sec:data}
The \Herschel\ Astrophysical Terahertz Large Area Survey (\hatlas; \citealp{Eales2010a}) was the largest-area submm survey conducted with the \Herschel\ Space Observatory, imaging around 590~deg$^2$ at 100 and 160\,\mum\ with PACS \citep{Poglitsch2010} and 250, 350 and 500\,\mum\ with SPIRE, as described by \citet{Ibar2010} and \citet{Pascale2010}.
We use data from Phase 1 covering 161~deg$^2$ in three equatorial fields at RA of approximately 9\hour, 12\hour\ and 14.5\hour, which benefit from extensive multi-wavelength coverage including the Sloan Digital Sky Survey (SDSS; \citealp{York2000}) and Galaxy and Mass Assembly survey (GAMA; \citealp{Driver2010}.
\Herschel-ATLAS sources were extracted from the Phase~1 PSF-filtered 250-\mum\ maps as described by \citet{Rigby2010} and Valiante et al. (in preparation), and are assumed to be point-like.
The source catalogues used for matching contain 122,862 sources detected at $\geq 5\sigma$ at 250\mum, corresponding to an average point-source detection limit of 27\,mJy. 
The optical catalogue was compiled as in \citet{Smith2011a} using all primary objects in SDSS DR7 \citep{Abazajian2009} with $r_\text{model}<22.4$, separating stars, quasars and galaxies using PSF/model magnitudes, spectroscopic redshifts and additional colour criteria. 
Spectroscopic redshifts exist for nine per cent of the $r<22.4$ sample, from the GAMAII redshift catalogue and other spectroscopic surveys \citep{Cannon2006,Abazajian2009,Croom2009,Jones2009,Ahn2013}, and all objects have photometric redshifts from SDSS DR7.
For the analysis in this paper we are interested in the 1,806,701 extragalactic objects in this catalogue. 
Further details of the optical catalogue and matching will be published in the counterpart identifications paper (Bourne et~al. in preparation).

\begin{figure}
\begin{center}
 \includegraphics[height=3.75cm]{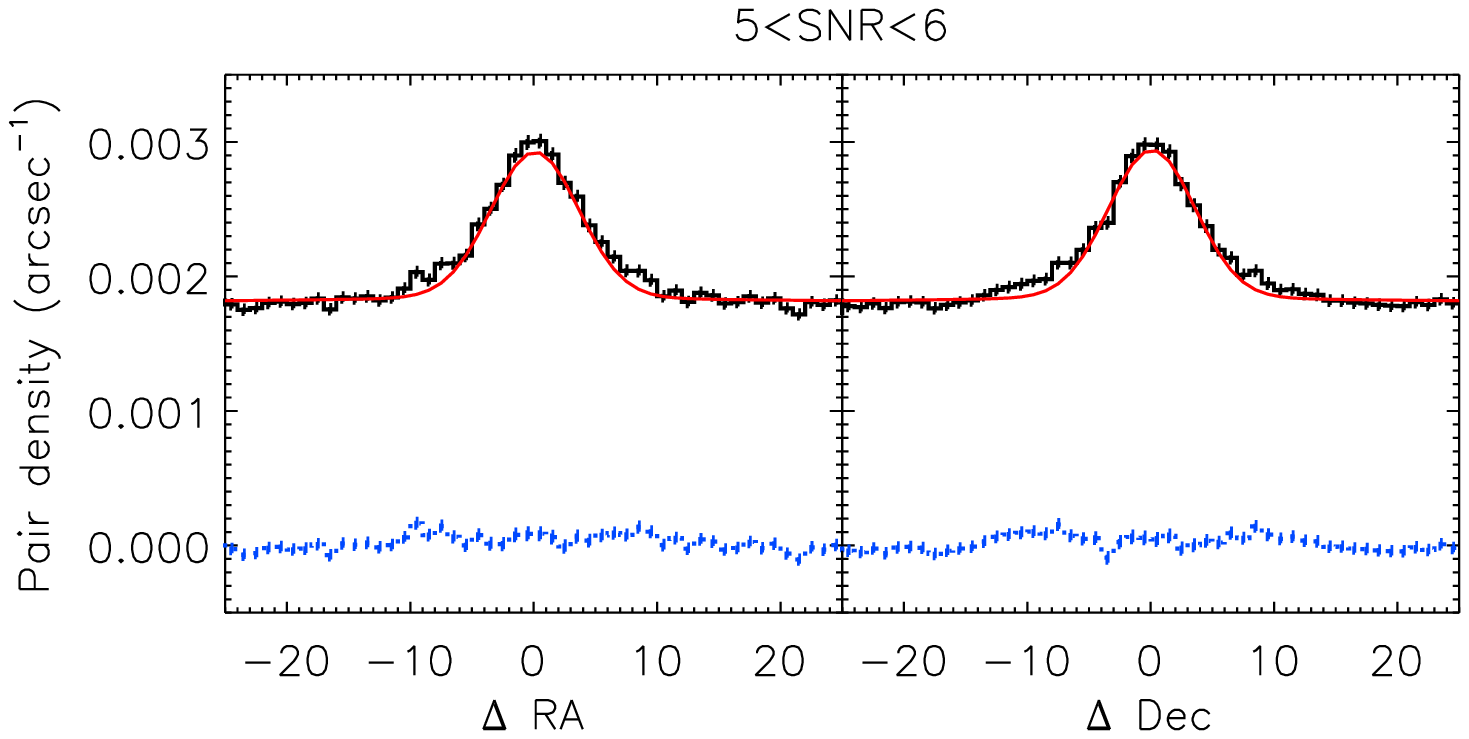}\\
\includegraphics[height=3.6cm,trim=0 0 0.75cm 0, clip]{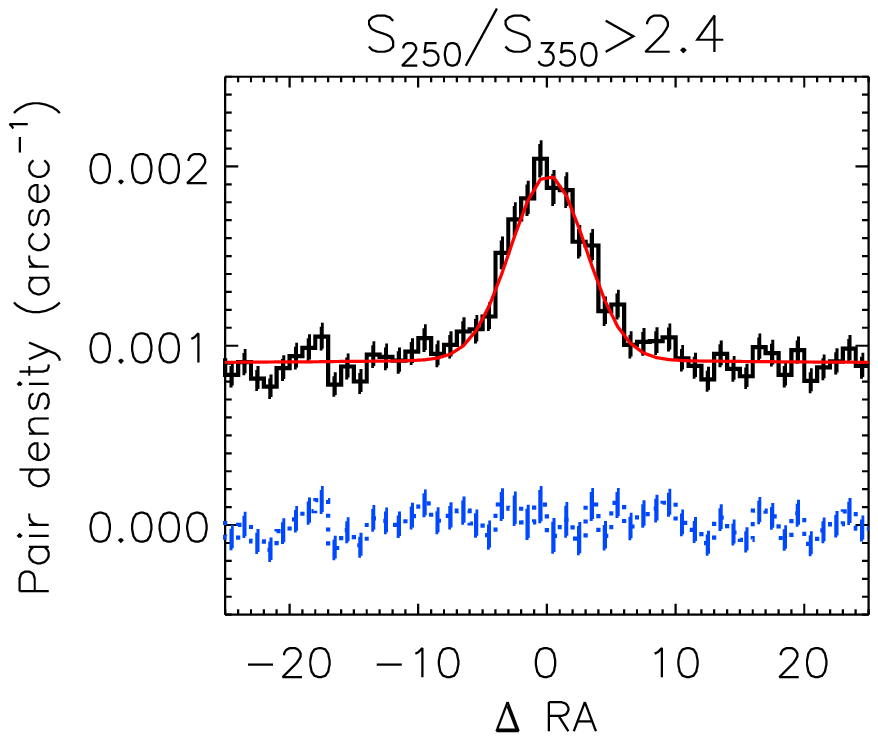} \hspace{-0.22cm}
\includegraphics[height=3.6cm,trim=2.65cm 0 0 0, clip]{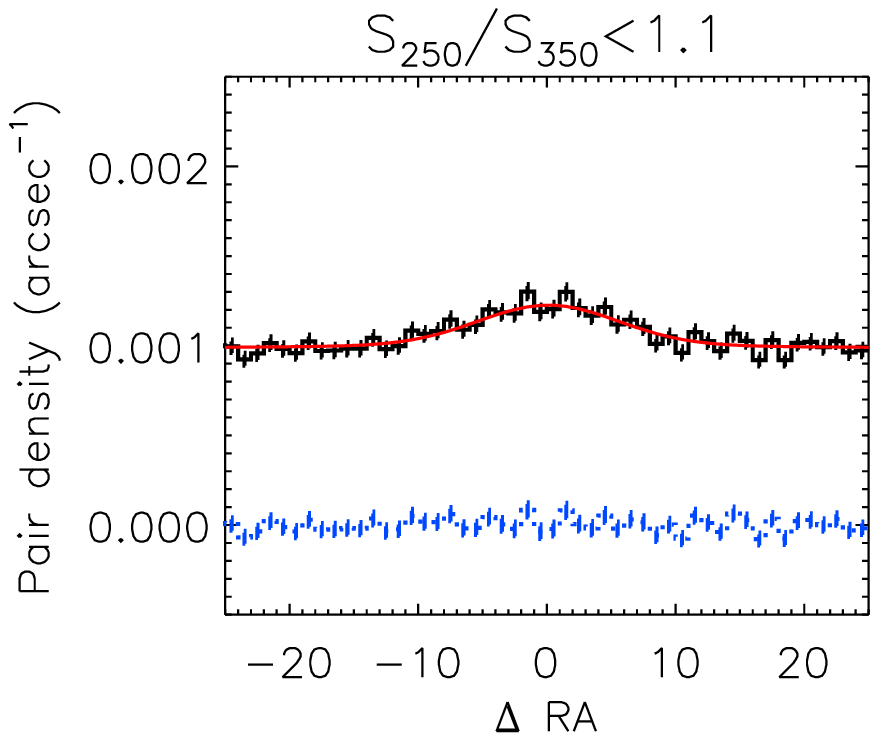}
\caption{Top: central 50 arcsec of the histograms of angular separations, in RA and Dec respectively, to all SDSS objects within a 100~arcsec square around each SPIRE source, averaged over all SPIRE sources in the $5<\text{SNR}<6$ bin.  The solid orange line shows the best-fitting model, consisting of the Gaussian positional errors of true counterparts, the power-law cross-correlation for correlated galaxies, and a constant background density. The dotted blue line shows the residual.
Bottom: the results (in RA only) in the same SNR bin, for the bluest SPIRE sources (left) and the reddest SPIRE sources (right) reveal a strong dependence of the width of the distribution on submm colour.
}
\label{fig:fithistogram} 
\end{center}
\end{figure}

We measure the separations of all SDSS galaxies around each SPIRE source up to a 50~arcsec separation in RA and Dec, thus forming a two-dimensional histogram of offsets between all SPIRE-SDSS pairs.
This distribution is assumed to consist of three components:
\begin{equation}
n(\Delta RA, \Delta Dec) = n_0+Q_0 f(r)+w(r)*f(r).
\label{eqn:fradec}
\end{equation}
These are (i) the constant background density of our SDSS sample $n_0=9.7\times10^{-4}$~arcsec$^{-2}$; 
(ii) the Gaussian distribution of radial offsets to true counterparts $f(r)$, representing the random positional errors, which is normalised by the fraction of true counterparts which are detected in SDSS ($Q_0$); and (iii) an additional contribution from correlated sources given by the cross-correlation $w(r)$ between SPIRE and SDSS positions, which is convolved with the positional error function.

If the width, $\sigma$, of $f(r)$ is simply the SPIRE positional error \sigmapos\ (neglecting the much smaller SDSS positional error), then measuring this width in $\Delta$RA and $\Delta$Dec should give results that follow the theoretical form 
\begin{equation}
\sigma_\text{th}(\text{SNR}) = 0.6 \dfrac{\text{FWHM}}{\text{SNR}}
\label{eqn:ivison2007}
\end{equation}
\citep{Ivison2007a}, where FWHM and SNR are the beam full-width at half-maximum and signal-to-noise ratio (respectively) in the 250-\mum\ image from which the sources were extracted (FWHM~$=18.1$~arcsec). 
This assumes an ideal situation where for example one submm source has one counterpart in the optical, but the real positional error \sigmapos\ can be increased by many effects including the presence of resolved sources, heavy confusion \citep{Hogg2001}, and clustering \citep{Chapin2011}.

Dividing the SPIRE catalogue into bins of 250-\mum\ SNR, we first estimate $w(r)$ using the modified \citet{Landy1993} estimator \citep{Herranz2001} to count pairs of data and random positions as a function of radial separation from 10--120~arcsec, and fit a power-law $w(r) = (r/r_0)^\delta$.
Fits to all SNR bins at these angular scales are consistent with an index $\delta=-0.7$ (within $\pm0.1$), equal to the index of the auto-correlation of SDSS galaxies \citep{Connolly2002,Smith2011a}. We therefore fix the index to this value and fit the correlation lengths $r_0$ (in arcsec) of $0.20\pm0.02$ at $5<\text{SNR}<6$, $0.20\pm0.03$ at $6<\text{SNR}<7$, $0.45\pm0.04$ at $7<\text{SNR}<9$, $0.57\pm0.06$ at $9<\text{SNR}<12$.
We fix $w(r)$ as measured in each SNR bin and fit the two-dimensional offset histograms on all scales $<50$~arcsec with the function in \eqnref{eqn:fradec},
thus obtaining measurements of $Q_0$ and \sigmapos\ in $\Delta$RA and $\Delta$Dec in each bin (see Fig.~\ref{fig:fithistogram} for example). 
These and all subsequent fits are obtained using Levenberg-Marquardt minimisation as implemented in the IDL package \textsc{mpfit} \citep{Markwardt2009}. Errors are measured from the covariance matrix of the parameters, and are scaled by the square-root of the chi-squared/degrees of freedom.

\begin{figure}
\begin{center}
 \includegraphics[width=0.42\textwidth]{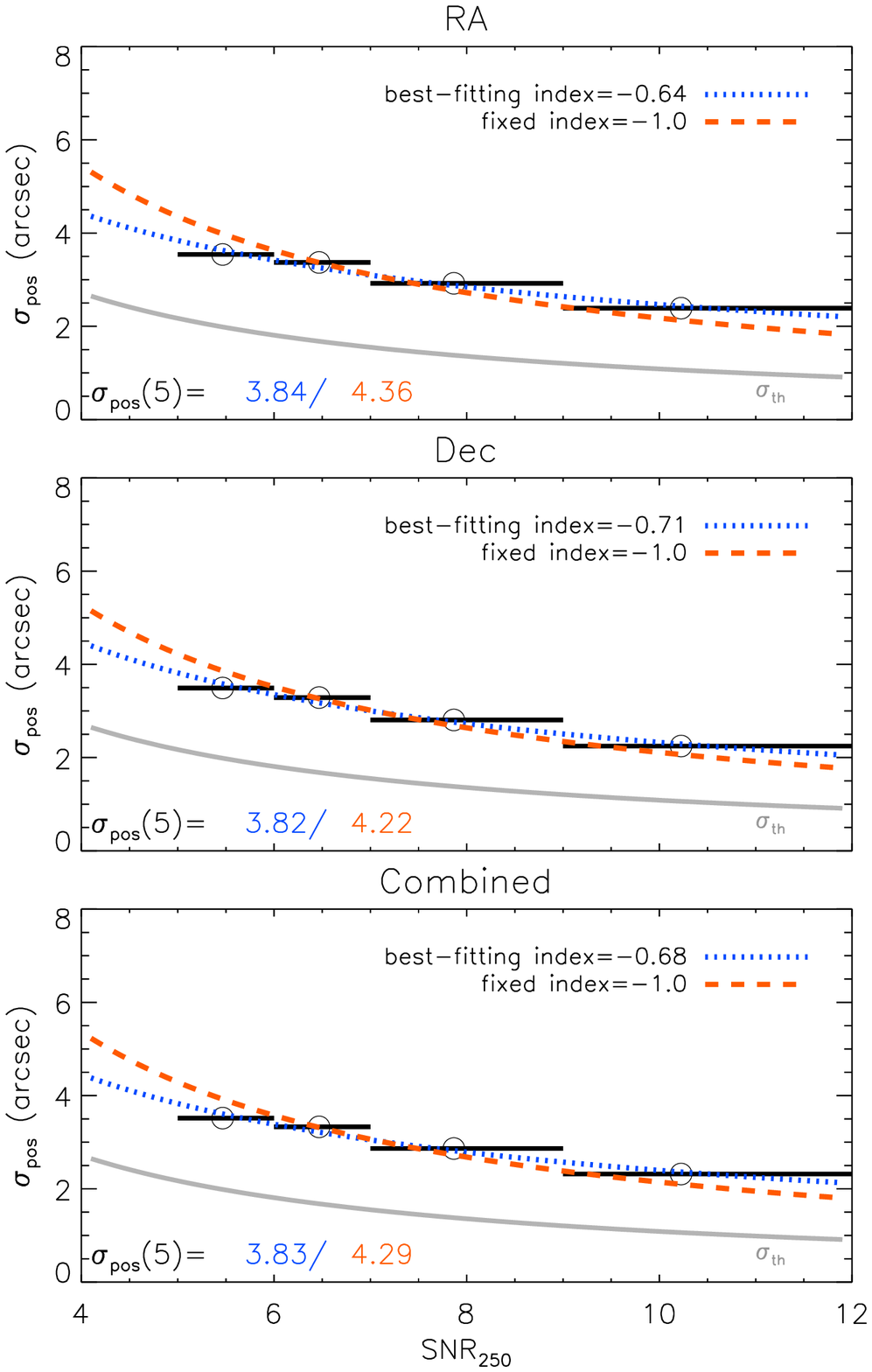}
\caption{Measured \sigmapos\ as a function of mean SNR in bins from the $\Delta$RA and $\Delta$Dec histograms; combined results given by the geometric mean of \sigmapos(RA) and \sigmapos(Dec). 
Best-fitting power-law models with free index (blue dotted line) or fixed index~$=-1$ (orange dashed line) are shown, and the value of \sigmapos(SNR=5) in the free/fixed-index models is printed in blue/orange. The grey line shows the theoretical \sigmapos\ from \eqnref{eqn:ivison2007}.} 
\label{fig:sigmapos_snr}
\end{center}
\end{figure}

We examine the behaviour of \sigmapos\ as a function of SNR (Fig.~\ref{fig:sigmapos_snr}) and fit a power law to estimate $\sigmapos(\text{SNR}=5)$.
In this way we measure the value of $\sigmapos(5)=3.83\pm0.02$ arcsec, with a best-fitting power-law index of $-0.68\pm0.01$. 
In comparison, \citet{Smith2011a} measured a value of $\sigmapos=2.40\pm0.09$ arcsec for all SPIRE sources with SNR~$>5$.
This value differs from our measurement mainly because \citeauthor{Smith2011a} did not bin by SNR, so the brighter sources would reduce the overall \sigmapos\ that they measured, and also because changes in the data reduction have reduced the noise levels measured in the Phase~1 maps: a 5-$\sigma$ detection corresponds to 27\,mJy in the catalogues used here, but was 32\,mJy in the catalogue used by \citeauthor{Smith2011a}

The expected value from \eqnref{eqn:ivison2007} is $\sigma_\text{th}(5)=2.17$ arcsec, a factor 1.8 smaller than our measurement, while the measured index of $-0.68$ is also significantly different from the expectation of $-1$ in \eqnref{eqn:ivison2007}, indicating the presence of factors other than SNR.

\section{Dependence on submm colour}
\label{sec:results}

\begin{figure}
\begin{center}
  \includegraphics[trim=0cm 10mm 0cm 0cm, clip=true, width=0.45\textwidth]{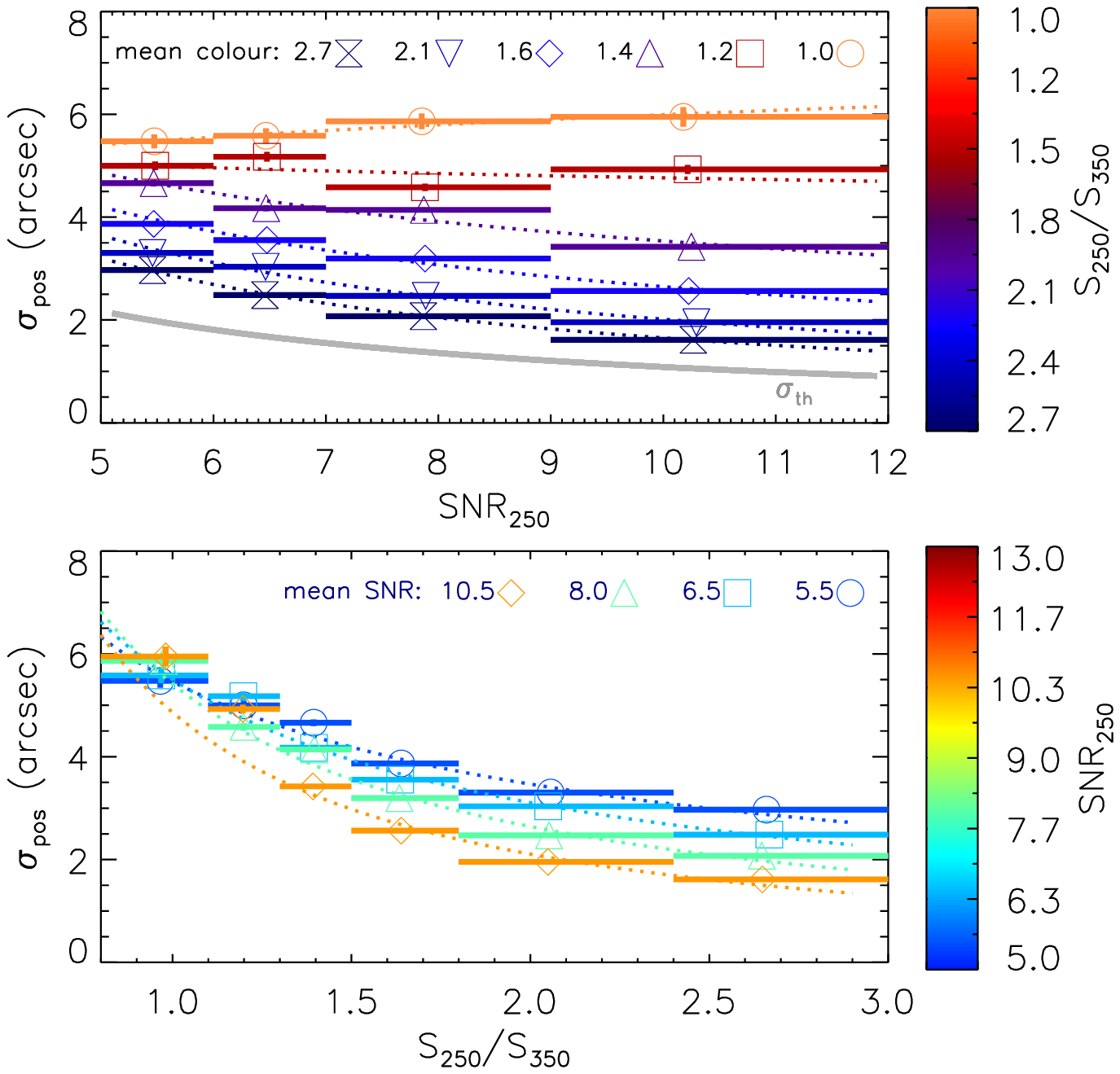}
 {\includegraphics[trim=0cm 0cm 0cm 5mm, clip=true, width=0.45\textwidth]{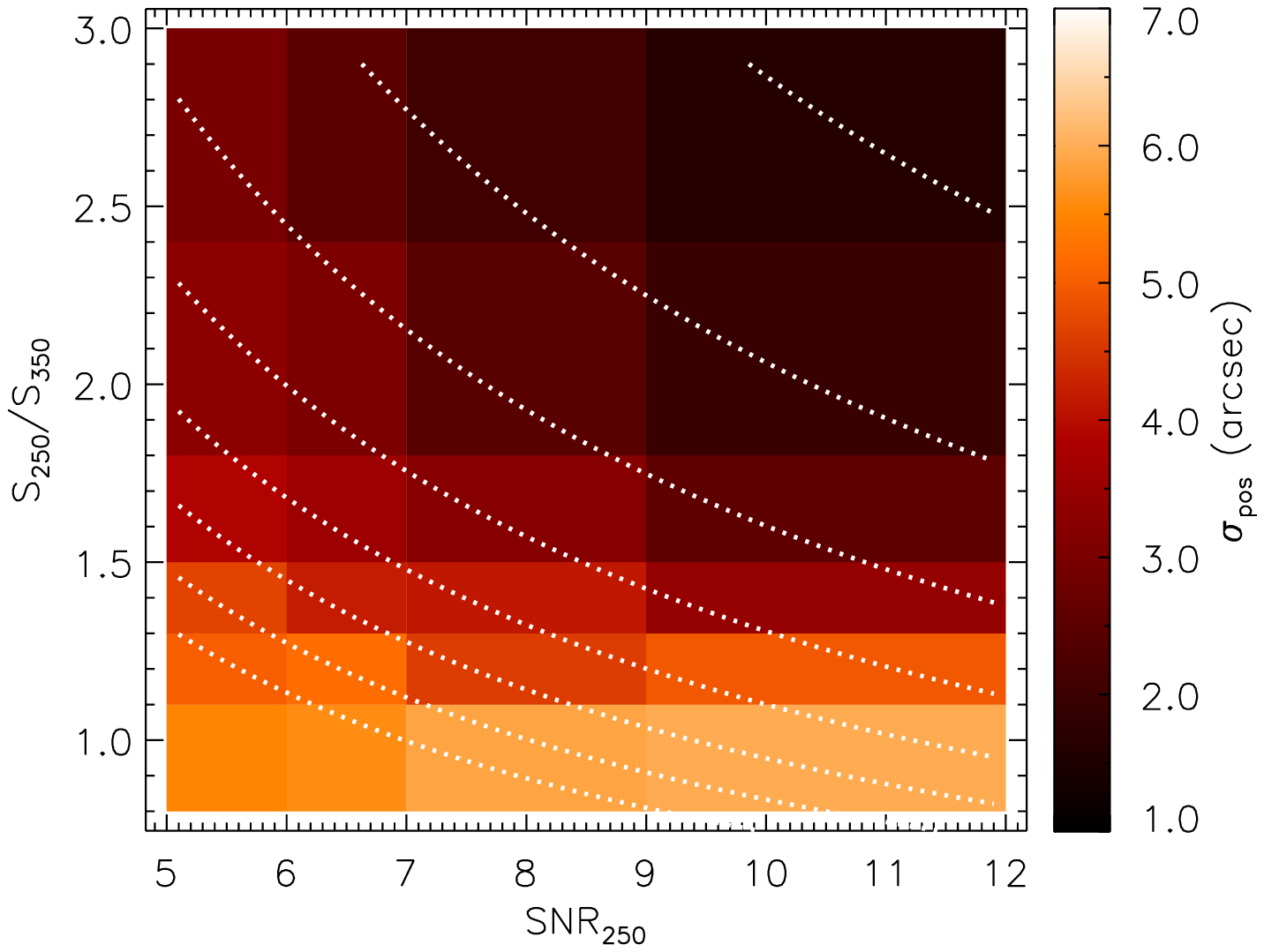}}

 \caption{Measured \sigmapos\ in SNR and colour bins, plotted as a function of mean SNR (top), mean colour (middle), and in the SNR/colour plane (bottom). 
 Symbols indicate the colour bin in the first panel, and the SNR bin in the second panel, and the mean colour or SNR of each is indicated in the legend.
 Dotted lines on the first two panels indicate the univariate power-law fits to each colour or SNR bin, and on the third panel indicate contours of the bivariate fit.
 The grey line shows the theoretical \sigmapos\ from \eqnref{eqn:ivison2007}.
   }
\label{fig:sigmapos_snr_col}
\end{center}
\end{figure}

\begin{table}
 \caption{Best-fitting power-law index and normalisation (at SNR=5) to \sigmapos(SNR) in each colour bin}
\begin{center}
\begin{tabular}{c c c c}
 \hline
 \multirow{2}{*}{$\frac{S_{250}}{S_{350}}$ bin} & \multirow{2}{*}{index} & $\sigmapos(5)$ & $\chi^2$\\
 & & (arcsec) & ($N_{\rm dof}$=2)\\
 \hline
 0.8--1.1 & $+0.15\pm0.06$ & $5.41\pm0.13$ & 0.4 \\
 1.1--1.3 & $-0.08\pm0.04$ & $5.03\pm0.08$ & 25.2 \\
 1.3--1.5 & $-0.46\pm0.03$ & $4.86\pm0.06$ & 25.5\\
 1.5--1.8 & $-0.67\pm0.02$ & $4.20\pm0.04$ & 23.5\\
 1.8--2.4 & $-0.86\pm0.01$ & $3.64\pm0.03$ & 29.7\\
 2.4--3.0 & $-0.96\pm0.02$ & $3.20\pm0.03$ & 2.2 \\
  \hline
\end{tabular}
\end{center}
\label{tab:sigmapos_snr_fits}
\end{table}

We can learn more about this problem by considering whether colour might be a factor in the apparent positional errors of submm sources.
In Fig.~\ref{fig:sigmapos_snr_col} we plot the measurements of \sigmapos\ obtained by dividing the sample into four bins of SNR and six bins of $S_{250} / S_{350}$ colour (examples of fits to two of these bins are shown in Fig.~\ref{fig:fithistogram}). 
These bins each contain between 900 and 8000 SPIRE sources (all but three of the bluest bins contain more than 2000).
Fitting power laws to these binned results as a function of both SNR and colour, we measure indices for $\sigmapos(\text{SNR})$ ranging from 
about zero in the reddest bins to $-1$ in the bluest (see Table~\ref{tab:sigmapos_snr_fits}).  
Likewise we fit $\sigmapos(S_{250}/S_{350})$ with indices ranging from $-0.66\pm0.02$ ($5<\text{SNR}<6$) to $-1.21\pm0.02$ ($9<\text{SNR}<12$), as shown in Fig.~\ref{fig:sigmapos_snr_col}. 
Evidently there is a smooth bivariate dependence of \sigmapos\ on both colour and SNR, which we can describe as
\begin{align}
 \sigmapos(S,C)=A\,  S^\alpha  C^\beta,   \label{eqn:fsnrcolor} \\
 S=\text{SNR}/5; C=\tfrac{S_{250}}{S_{350}}/3. \nonumber
\end{align}
The best-fitting parameters are $A=\sigmapos(1,1)=2.54\pm0.12$~arcsec, $\alpha=-0.77\pm0.07$ and $\beta=-0.93\pm0.06$ ($\chi^2=1243$, degrees of freedom $N_{\rm dof}=21$). 
The high $\chi^2$ indicates that the fit is not perfect, but this is not unexpected since we do not necessarily expect a power-law dependence on colour.

Hence for the bluest 5-$\sigma$ sources $\sigmapos=2.54$ arcsec, which is close to the prediction from \eqnref{eqn:ivison2007} of 2.17 arcsec, suggesting that the histogram of offsets for blue sources is well described by the positional errors of true counterparts plus the cross-correlation function. 
Furthermore, the width of this histogram for blue sources more closely follows the predicted SNR$^{-1}$ dependence given by \eqnref{eqn:ivison2007}, but that dependence weakens or vanishes for redder SPIRE sources. We note that some deviation from \eqnref{eqn:ivison2007} may be expected from the effects of instrumentation and map-making, but the dependence on the observed submm colour in sources extracted from 250-\mum\ positions cannot be explained in this way.

We tested whether the results could be caused by confusion alone using a simulation in which additional sources were added into a 250-\mum\ map at random positions, and then extracted in the same way as the catalogues (see Valiante et al. in prep. for details of the simulation). For the simulated sources we found $\sigmapos(5)=2.3\pm0.1$ arcsec and $\alpha=-0.7\pm0.1$, but no difference was found between red and blue colour bins.
We also repeated the analysis with catalogues based on a matched filter extraction \citep{Chapin2011}, which minimises the confusion that affects PSF filtering, and found that \sigmapos\ for real sources was smaller by a factor of $0.92\pm0.04$ (based on the fit for $A$), but followed similar trends with SNR and colour. 

Positional errors in point-source catalogues can be increased if some sources are resolved, although this would be more common among blue rather than red sources, assuming that colour indicates redshift. We excluded this possibility by repeating the analysis using only SDSS galaxies with (spectroscopic or photometric) $z>0.2$, and still found a strong colour dependence.

\section{Astrophysical Interpretations} 
\label{sec:discussion}

The increased positional offsets between red SPIRE sources and their SDSS associations must be attributed to a phenomenon that is widespread among galaxy counterparts (in order to affect the overall statistics so strongly), and whose effect increases continuously towards redder and brighter submm sources.
It may be a phenomenon that increases the positional errors of red SMGs preferentially, or that acts on a certain population to boost their fluxes at longer wavelengths and simultaneously alter their positions, or increases the chance of alignment between red SMGs and another galaxy population.

The ubiquitous blending which occurs in single-dish submm images could play a role here.
For example, galaxies that are more clustered will suffer the most from blending and will be flux-boosted at longer wavelengths (due to coarser resolution), making them appear redder. If the clustered sources are sufficiently blended they will also suffer greater positional uncertainties due to the movement of the centroid when multiple point sources are superimposed.
The implication of this is that SMGs of different colours may have differing positional uncertainties as a result of confusion, which must be taken into account in the probabilities assigned to potential counterparts.
This scenario may be supported by recent work suggesting that many high-redshift SMGs in single-dish surveys could be blends  \citep[although see also \citealt{Karim2013,Koprowski2013,Targett2013}]{Ivison2002,Ivison2007a,Hayward2012a,Hodge2013}.
However, we note that the redshift distribution of the wide and shallow \hatlas\ is very different to that of most submm surveys, and the average number of sources per beam will be smaller because more sources are at low redshifts ($z\lesssim0.5$).

An alternative hypothesis is that gravitational lensing can also affect the offset distribution. Differential k--corrections in the optical and submm mean that an apparent optical counterpart is likely to be a foreground lens, while the submm survey detects only the lensed image of the background source, which is offset from the lens itself. The lensed image may be visible in a deep optical/near-IR image but is likely to be too faint for SDSS. Lensing does not alter the colour of the source, but the lensing optical depth increases with source redshift and will therefore be increased for redder submm samples (submm colour is a rough indicator of redshift, e.g. \citealp{Amblard2010}).
Furthermore, the magnification of the lensed source means that this effect is more likely to be seen in submm sources with higher fluxes.
\citet{Lapi2012} predict number counts of lenses that make up a significant fraction of the total number counts at 350\,\mum, especially for $\log(S_{350})\gtrsim1.5$ (see also \citealt{Negrello2007,Lapi2011}), which is supported by observational evidence in the \hatlas-selected lens sample of \citet{Gonzalez-Nuevo2012}.
Several observational studies have measured significant lensing-induced cross-correlation signals between high-redshift SMGs and low-redshift galaxy populations \citep[although \citealt{Blake2006} report a non-detection]{Almaini2005,Wang2011a,Gonzalez-Nuevo2013}, 
as well as boosting of submm source counts resulting from lensing by foreground large-scale structure \citep{Aretxaga2011,Noble2011}.
If this is the explanation for our results, then the positional errors are not dependent on colour, but a large fraction of \Herschel\ sources are weakly lensed and are likely to be mis-identified with the lens when the true optical counterpart is below the limiting magnitude \citep[consistent with the conclusions of][]{Gonzalez-Nuevo2012}.

\begin{figure*}
\begin{center}
 \includegraphics[width=\textwidth]{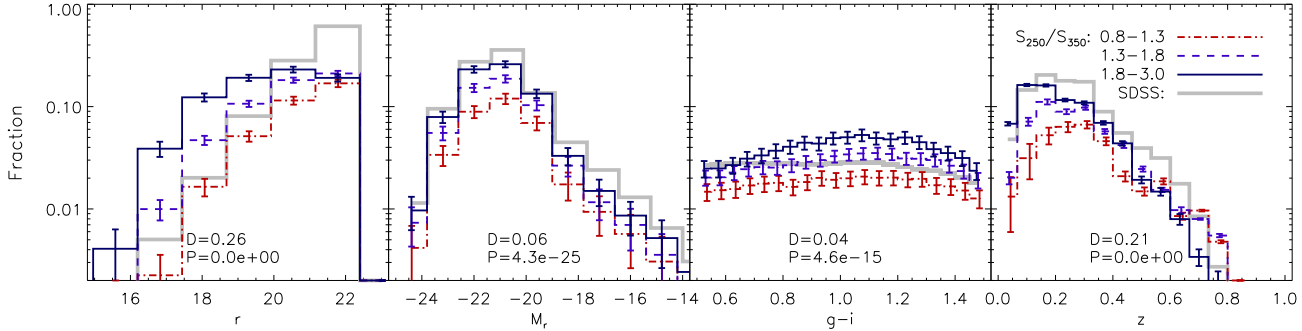}
\caption{Distributions of optical properties and redshifts of the SDSS counterparts to SPIRE sources, as a fraction of the total number of SPIRE sources in each of three submm colour bins. The background SDSS distributions are shown in grey. Absolute magnitudes $M_r$ and $g-i$ colours are k-corrected to the rest frame. In each panel we show the Kolmogorov-Smirnov statistic ($D$) computed from the bluest and reddest bin, and the associated probability ($P$) that the two samples are drawn from the same distribution.}
\label{fig:qm_colour}
\end{center}
\end{figure*}

\subsection{Lensing or Clustering?}

We can try to discriminate between these two scenarios by looking for differences between the populations of counterparts to red and blue \hatlas\ sources. 
We estimate the distributions of $r$-band magnitude, absolute magnitude $M_r$, redshift and $g-i$ colour for `true' counterparts to \hatlas\ sources.
To estimate rest-frame $M_r$ and $g-i$ we apply k-corrections by fitting the $gri$ photometry using {\sc kcorrect} v4.2 \citep{Blanton2007}, with the redshifts described in Section~\ref{sec:data}, and measuring the distributions from the 80 per cent of fits which have $\chi^2<10$.
We determine the distribution of each quantity ($r$, $M_r$, etc) for all SDSS sources within 10~arcsec of all \hatlas\ sources, and subtract the `background' distribution of SDSS sources within 10~arcsec of the same number of random positions in the same fields. We normalise this by $Q_0$ from \eqnref{eqn:fradec}, i.e. the estimated fraction of \hatlas\ sources in that bin which have a counterpart above the $r$-band limit of the catalogue \citep[see][]{Smith2011a}. Note that this method for sampling the population of counterparts removes random associations only, and can still be contaminated by foreground lenses and correlated objects.

Figure~\ref{fig:qm_colour} shows the distributions of optical properties estimated in this way, for \hatlas\ sources in bins of submm colour. 
The results of a Kolmogorov-Smirnov test (shown on the figure) indicate that the distributions of blue and red submm sources are significantly different.
The $r$-band and redshift distributions show that the optical counterparts to red submm sources are fewer, fainter and at higher redshifts compared to those of blue submm sources, but have no strong predisposition with respect to optical colour.  
These results are to be expected if submm colour is primarily dependent on redshift.
If submm-red galaxies are either more strongly clustered or more likely to be lensed by a galaxy then we might expect their optical associations to have redder optical colours and brighter absolute magnitudes, since these properties are associated both with increased clustering \citep{Zehavi2010} and with (strong) lenses \citep{Auger2009}.
The results show no evidence for such tendencies, but we note that lensing by galaxy groups would be consistent with this null result.

\begin{table}
\begin{center}
\caption{Statistics of matching candidates in SDSS within 10~arcsec of SPIRE sources in bins of SNR and colour}
\begin{tabular}{c c c c c}
\hline
& \multicolumn{4}{c}{SNR bin}\\
$\frac{S_{250}}{S_{350}}$ bin & 5--6 & 6--7 & 7--9 & 9--12 \\
\hline
                   & \multicolumn{4}{c}{average no. matches in 10 arcsec} \\
{0.8--1.3} & 1.22 & 1.25 & 1.26 & 1.28 \\
{1.3--1.8} & 1.34 & 1.39 & 1.43 & 1.48 \\
{1.8--3.0} & 1.42 & 1.50 & 1.57 & 1.61 \\
\hline
                   & \multicolumn{4}{c}{fraction with $>1$ match} \\
{0.8--1.3} & 0.17 & 0.20 & 0.20 & 0.21 \\
{0.3--1.8} & 0.26 & 0.29 & 0.31 & 0.34 \\
{1.8--3.0} & 0.32 & 0.36 & 0.41 & 0.43 \\
\hline
\end{tabular}
\label{tab:multiplicity}
\end{center}
\end{table}

\begin{figure}
\begin{center}
 \includegraphics[trim=0cm 10mm 0cm 0cm, clip=true, width=0.45\textwidth]{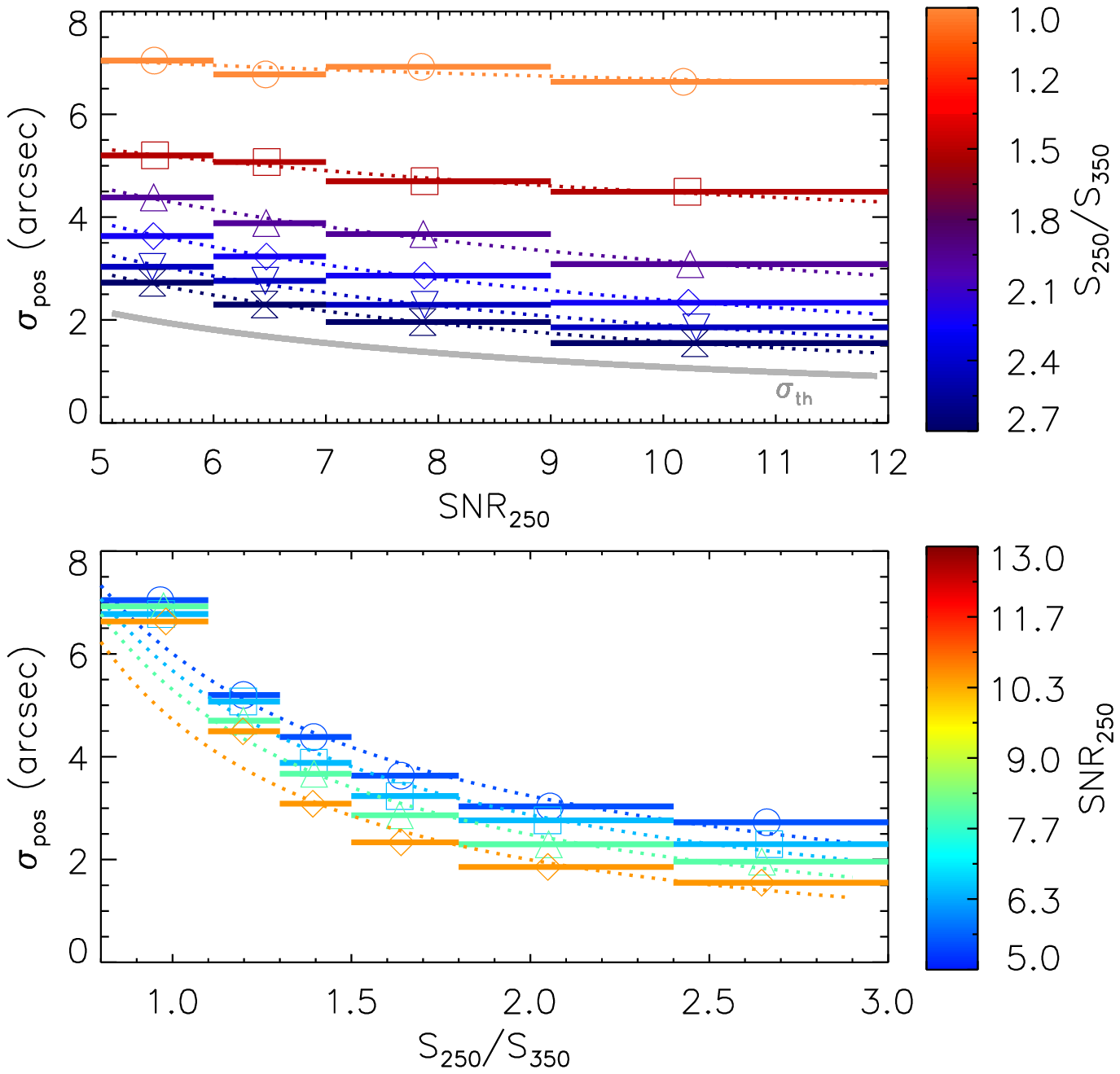}
 {\includegraphics[trim=0cm 0cm 0cm 5mm, clip=true, width=0.45\textwidth]{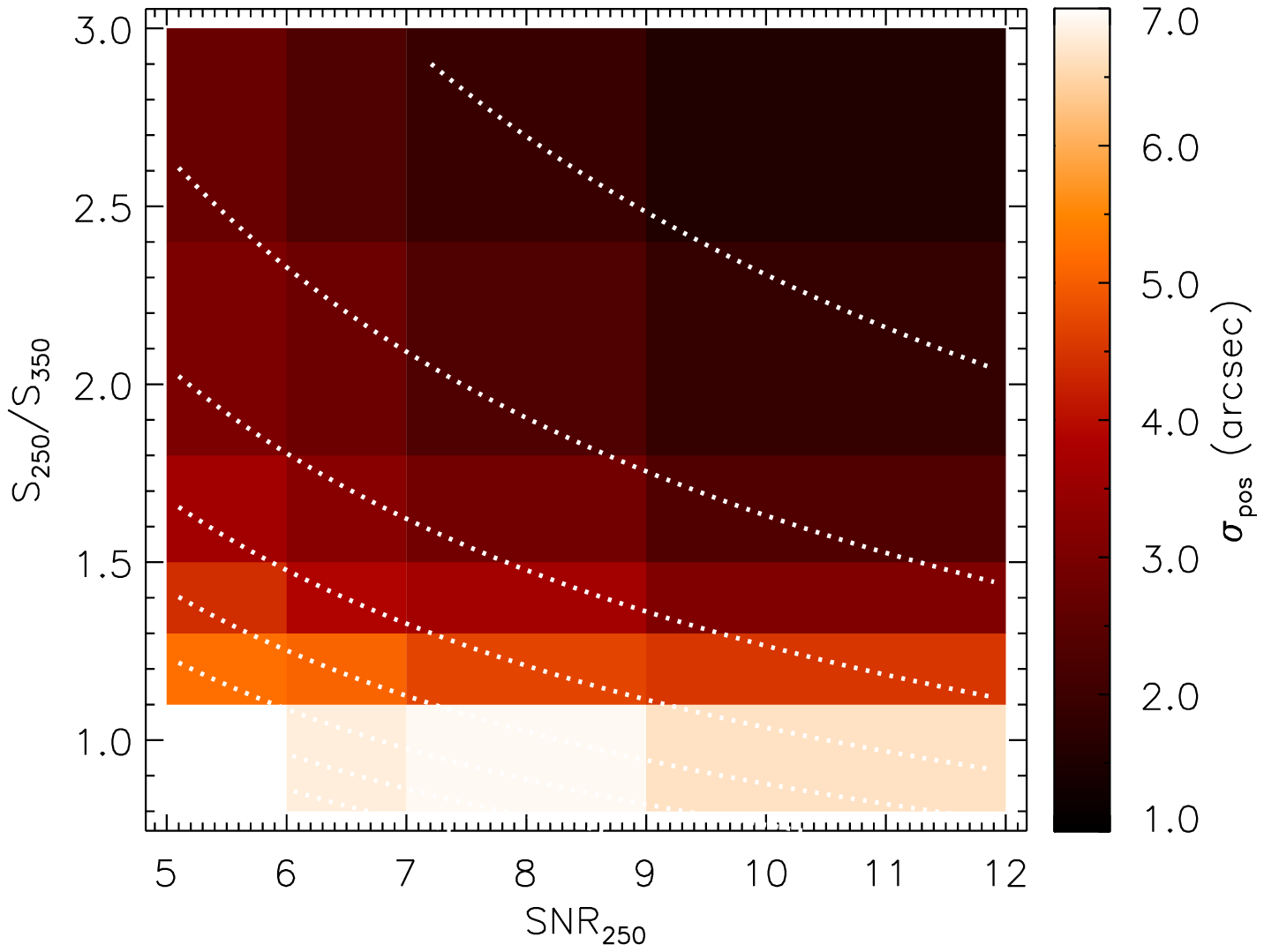}}

 \caption{Measured \sigmapos\ in SNR and colour bins, as in Fig.~\ref{fig:sigmapos_snr_col}, but for the nearest SDSS galaxy to each SPIRE source.}
 \label{fig:sigmapos_nearest}
\end{center}
\end{figure}

Another possible discriminator between the clustering and lensing explanations is in the multiplicity of potential counterparts. 
We find that submm-red \hatlas\ sources on average have fewer SDSS candidate matches within a 10~arcsec radius than blue ones in a given SNR bin, and are about half as likely to have more than one potential counterpart within that radius (Table~\ref{tab:multiplicity}). 
This radius corresponds to around 20\,kpc at $z=0.1$, and 65\,kpc at $z=0.5$.
This result reinforces the interpretation that the offset distribution is not broadened by correlated neighbours, since these would increase the number of candidate matches (although we have only accounted for correlated neighbours within the SDSS magnitude limit).
We further tested this possibility by repeating the analysis using only the nearest SDSS counterpart to each SPIRE source, and found results (shown in Fig.~\ref{fig:sigmapos_nearest}) similar to those from all SDSS-SPIRE pairs: $A=2.24\pm0.12$ arcsec, $\alpha=-0.73\pm0.07$, $\beta=-1.05\pm0.06$ ($\chi^2/N_{\rm dof}=3821/21$). The results are therefore not simply caused by a failure to account for the correlation function, since even the nearest counterparts on the sky have greater offsets for red submm sources.

A redshift cut could provide a test of the lensing hypothesis, since lensing magnification is greatest when the lens is halfway between the source and observer in angular-diameter distance. For this reason, and due to the redshift distribution of SMGs, strong lenses (with magnifications greater than 2) are exceptionally rare at $z<0.1$. Our results are more likely to be affected by weak lensing with lower magnifications, since this is much more common, but weak lensing will also be correspondingly reduced at the lowest redshifts. 
We divided our SDSS sample into two redshift bins to test whether the colour dependence is equally strong in SDSS associations at low redshifts.
We used the spectroscopic and photometric redshifts described in Section~\ref{sec:data}. 
In order to avoid bias in the fluxes and colours of resolved SPIRE sources at low redshifts we substituted the extended flux measurements for resolved sources \citep[as in][]{Rigby2010} in place of the PSF-filtered measurements used hitherto. 
Repeating the \sigmapos(SNR,colour) analysis with all SPIRE sources, and with 76,460 and 1,734,476 SDSS sources at $z<0.1$ and $z>0.1$ respectively, we found a lesser colour dependence but stronger SNR dependence in the best fit to \eqnref{eqn:fsnrcolor}: 
$A=2.75\pm0.21$ arcsec, $\alpha=-1.31\pm0.12$, $\beta=-0.43\pm0.09$ in the low redshift sample; and $A=2.99\pm0.10$ arcsec, $\alpha=-1.06\pm0.05$, $\beta=-0.36\pm0.04$ in the high redshift sample 
(compared with $A=2.54\pm0.12$ arcsec, $\alpha=-0.77\pm0.07$, $\beta=-0.93\pm0.06$ as measured in the full sample).
The fact that reduced colour dependence is seen in each of these bins relative to that in the full sample suggests that the colour dependence is at least partially caused by different \sigmapos\ behaviour at different redshifts. 
There is, however, no evidence that the colour dependence is weaker at $z<0.1$ than $z>0.1$, but the measurement of \sigmapos\ may be sensitive to different effects in the various colour and redshift bins, and crucially we do not know the redshifts of the SPIRE sources themselves. 
For example, blue SPIRE sources and low-redshift SDSS galaxies are likely to be cospatial and associations will be dominated by true counterparts and galaxy clustering. 
Conversely, it is unlikely that many red SPIRE sources (especially with $S_{250}/S_{350}\lesssim1$) are at $z<0.1$, and their SDSS associations in this redshift range could be dominated by lensing structures. In that case the measured \sigmapos\ probes fundamentally different mechanisms for blue and red SPIRE sources, and the colour dependence of that measurement is hard to predict.
Finally, we note that SPIRE sources at low redshifts will suffer from increased positional errors due to being (at least marginally) resolved, and the errors on their fluxes (and colours) are also increased relative to point sources. 
This means that our measurements probably over-estimate the positional errors of point sources, and the SNR and colour dependence could be affected by all of these uncertainties.

\subsection{A Comparison with HerMES}
\label{sec:hermes}
If lensing is responsible for the broadening of the offset distributions then we should see a greater number of associations to the red SPIRE sources than if we were only detecting true counterparts.
We can explore this possibility by comparing the fraction of \hatlas\ sources with a counterpart in SDSS with the fraction that would be obtained in HerMES \citep{Oliver2012} using the same selection criteria. This is instructive since HerMES counterparts are much less likely to suffer from false identifications with lenses due to the use of 24-\mum\ priors at 3.6-\mum\ positions, which have positional errors $\sim1$~arcsec \citep{Roseboom2010}. 
We used HerMES catalogues from HeDaM,\footnote{\url{http://hedam.lam.fr/HerMES/index/survey}} which contain a total of 13760 sources with $S_{250}>27.1$\,mJy (our average flux limit) in the EGS and UDS fields. We matched these with deep ($r_{AB}<25$) optical catalogues from the CFHTLS wide survey,\footnote{\url{http://terapix.iap.fr/rubrique.php?id_rubrique=268}}, using a 3-arcsec search radius, and found 12840 matches (93 per cent) in total, and 4960 (36 per cent) with $r_{AB}<22.4$ (the SDSS limit used for \hatlas\ matching).
To replicate the red sample, 5896 of the HerMES sources have $S_{250}/S_{350}<1.3$, of which 781 (13 per cent) have a match with $r_{AB}<22.4$.
These percentages differ by up to three per cent between the EGS and UDS fields as a result of cosmic variance.
In comparison, the best fitting normalisations ($Q_0$) of the offset distributions from Section~\ref{sec:results} indicate that 43 per cent of red sources in \hatlas\ have a galaxy counterpart at $r<22.4$ in SDSS.
A more rigorous analysis of the fraction of \hatlas\ sources with SDSS galaxy counterparts following the method of \citet{Fleuren2012} yields a value of $33\pm2$ per cent for the same red sample, and $51.4\pm0.1$ per cent for all SPIRE sources (this estimate avoids bias from clustered and multiple counterparts; see Bourne et~al. in preparation). 
These would appear to be boosted by false counterparts if the corresponding fractions in HerMES are 13 per cent of red sources and 36 per cent of all sources ($S_{250}>27.1$\,mJy).

Similar conclusions can be drawn from the redshift distribution of the HerMES sources in these samples: the photo-z catalogue from CFHTLS \citep{Ilbert2006,Coupon2009} has reliable redshifts ($0<z<6$) for 67 per cent of the sample (those with $i_{AB}<24$), of which 35 per cent overall (and 15 per cent of the red sample) are at $z<0.5$. A smaller but more complete redshift sample is provided by the deep CANDELS photo-z catalogue of the EGS \citep{Dahlen2013},\footnote{\url{https://rainbowx.fis.ucm.es/Rainbow_navigator_public/}} with redshifts for all 457 matches in the overlap region. In this subset 107 ($23\pm2$ per cent) of all SPIRE sources with $S_{250}>27.1$\,mJy are at $z<0.5$, but only 21 ($10\pm2$ per cent) of the 213 red SPIRE sources are at $z<0.5$. 
The errors due to cosmic variance are about 20 per cent of these numbers, using the Cosmic Variance Calculator \citep{Trenti2008}.\footnote{\url{http://casa.colorado.edu/~trenti/CosmicVariance.html}}
The redshift histograms in Fig.~\ref{fig:qm_colour} suggest that significantly higher fractions (49 per cent of all, and 31 per cent of red sources) in \hatlas\ are at $z<0.5$.
In summary, the comparison with flux-matched samples in HerMES suggests that the apparent fraction of \hatlas\ sources with low-redshift, SDSS-detected counterparts is considerably boosted by false associations, which are likely to be at lower redshifts than the sources themselves.

\subsection{Lensing Simulations}
We can further explore the plausibility of a lensing bias in the positional offsets with simulations of SMG lensing. 
\citet{Gonzalez-Nuevo2013} simulated the effects of weak lensing (magnifications $<2$) by galaxies and large-scale structure on the cross-correlation between high-redshift sources in \hatlas\ (Phase~1) and low-redshift galaxies in SDSS. They selected a SPIRE sample with $S_{250}>35$\,mJy, SNR$_{350}>3$, and no SDSS counterpart with reliability $>0.8$, thus sampling redshifts $\gtrsim1.5$ \citep[see][]{Lapi2011}; 
and a sample of SDSS galaxies with $r<22$ and photometric redshifts $0.2\leq z\leq0.6$.
They ran three sets of simulations, each using the SDSS sample as the foreground population and randomising the positions of the SPIRE background sample in 1920 Monte-Carlo realisations. In the first set of simulations they modelled the effects of weak lensing by galaxy-sized single-isothermal-sphere (SIS) haloes; in the second they replaced these with cluster-sized Navarro-Frenk-White (NFW) haloes; and in the third they included no lensing. We use the difference between the simulations with lensing and the one without, to determine the lensing-induced excess of SDSS objects in each radial bin around the average high-redshift SPIRE source. 

\begin{figure*}
\begin{center}
 \includegraphics[width=0.65\textwidth]{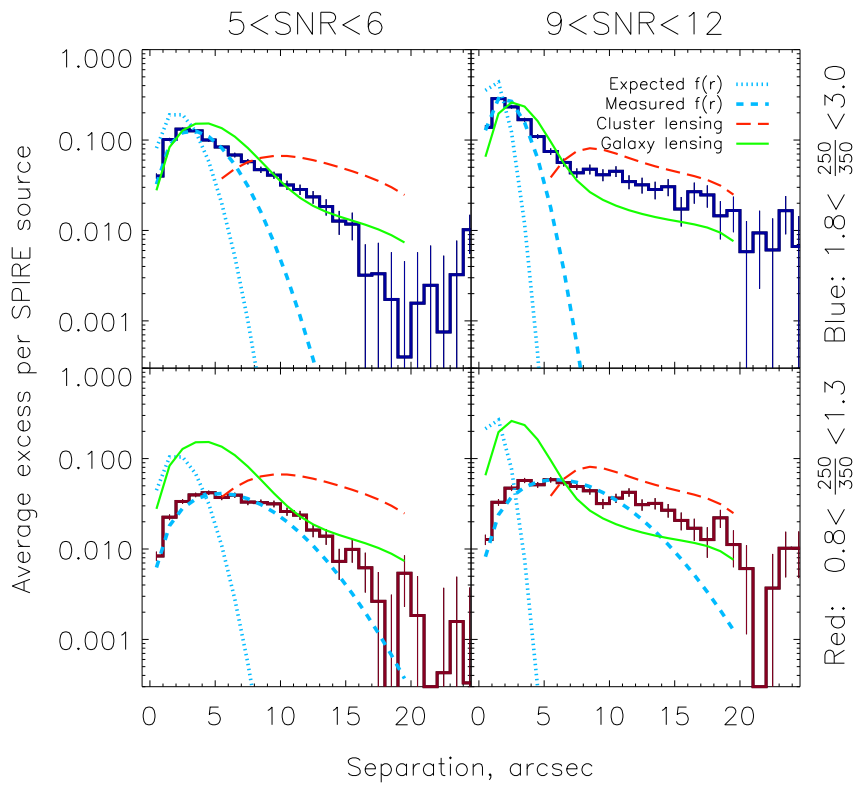}
 \caption{Histograms of the average number of SDSS galaxies in radial bins around each SPIRE source, for two bins of SNR and two of $S_{250}/S_{350}$ colour. The background density has been measured in random positions and subtracted from these to give the excess. Dotted blue line: the expected distribution of offsets of true counterparts according to \eqnref{eqn:ivison2007}. Dashed blue line: the distribution of offsets given by the best-fitting Gaussian $f(r)$ in each bin, as shown in Fig.~\ref{fig:sigmapos_snr_col}. Note that the histogram deviates from the theoretical expectation (dotted line) especially in the red bins, and also from the best fit (dashed line) at large separations, suggesting an additional contribution. 
The green solid (orange dashed) lines show the profile of offsets to low-redshift lenses around high-redshift 350-\mum\ sources in the simulations of galaxy (cluster) lenses. The normalisation of these lines has been rescaled to provide an upper limit since the simulation flux limit is not directly comparable to the observed data. Note that the simulations of cluster lenses (orange lines) are truncated at 6 arcsec and the inner profile is therefore potentially underestimated.
 } 
\label{fig:simfr}
\end{center}
\end{figure*}

In Fig.~\ref{fig:simfr} we show our observed histograms of radial offsets (integrated azimuthally) from each SPIRE source to all SDSS galaxies, averaged over all SPIRE sources in two bins of colour and two of SNR. For comparison, the best-fitting Gaussian positional error functions of counterparts described in Section~\ref{sec:results} and the theoretical ones given by \eqnref{eqn:ivison2007} are plotted as blue dashed and dotted lines respectively (the normalisation of both of these matches the normalisation of the fits in Fig.~\ref{fig:sigmapos_snr_col}).
The predictions from the simulations (plotted in green and orange) indicate a broad distribution of offsets for the lensing excess, due to either galaxy-sized or cluster-sized haloes respectively. 
The simulated offset distributions are calculated per SPIRE source and are the same in each bin, but to simulate the effect of positional errors in extracted sources they have been convolved with a Gaussian whose width is the measured \sigmapos\ of blue SPIRE sources in the same SNR bin (assuming that the blue bin contains negligible broadening from lensing).
The normalisation of the predicted offset distribution depends on the SPIRE flux limit, although the radial form does not change.
The simulated sources are extracted at 350\,\mum\ with a range of flux limits but these are not directly comparable to the observations of sources extracted at 250\,\mum, since the fraction of 250-\mum\ sources which are lensed will be lower than that of 350-\mum\ sources of the same flux \citep{Lapi2011}.
Furthermore the observations are not limited to any redshift range, while the simulations are limited to SPIRE redshifts $\gtrsim1.5$ and SDSS redshifts $0.2\leq z\leq0.6$. 
We therefore show the radial form of the predictions in Fig.~\ref{fig:simfr}, but the normalisation of each line is rescaled so that the total excess associations from the lensing simulation is equal to the sum of the measured SDSS offset histogram for bright ($9<\text{SNR}<12$), red ($0.8<S_\text{250}/S_\text{350}<1.3$) submm sources (i.e. the histogram in the lower-right panel of Fig.~\ref{fig:simfr}).
This provides a maximal upper limit on the contribution from lensing-induced offsets at 250\,\mum, under the assumption of no true counterparts or direct physical correlation with SDSS galaxies.

The SPIRE colour bins might be considered to sample different redshift ranges, and so we would expect the greatest lensing contribution in the red (highest-redshift) bin.  The simulated upper-limit lensing contributions are shown in all bins for comparison, but we expect the true lensing contribution to be much lower in the blue bin since it is dominated by lower-redshift sources which are less likely to be lensed and more likely to have an SDSS counterpart.

We can see that the offsets for blue SPIRE sources are indeed well described by a reasonably narrow Gaussian at small scales, although it is apparent that this is broader than the theoretical expectation. The profile at larger radii will include physical correlations with other galaxies at the same redshifts as the SPIRE sources, although it may also be biased by lensing structures.
For red SPIRE sources we see that the best-fitting Gaussian is biased to a much broader width than expected, and the profile is likely to be a combination of a relatively small number of true counterparts at small radii (with offsets similar to the counterparts to blue SPIRE sources) and lensing structures at larger radii.
The profile expected from galaxy-cluster lensing at radii $\gtrsim8$~arcsec is consistent with the excess in associations to SPIRE sources, and the upper limit is consistent with the data. The predictions for galaxy-galaxy lensing peak at smaller radii and could be responsible for associations at separations $\lesssim5$~arcsec. Intermediate-sized haloes (i.e. groups) could also introduce correlations at scales intermediate between those of galaxies and clusters. 

The results show that in principle it is quite possible for lensing-induced `matches' to contribute significantly to the offset histograms of candidate matches to red SPIRE sources, and that galaxy-galaxy lensing should be important only at small angular separations, but large-scale-structure lensing will dominate at larger separations.
The alternative explanation is that the associations at large separations are due to physical correlations between SPIRE and SDSS sources at the same redshifts, but in that case the red SPIRE sources must be at low redshift and should have enough counterparts in SDSS for the $f(r)$ term to be visible above the larger-scale cross-correlation, as it is in the histograms for blue SPIRE sources.

\section{Conclusions}
We have shown that \Herschel-SPIRE sources of different submm colours have different distributions of positional offsets from their SDSS associations. We demonstrate a smooth dependence of these offsets on submm SNR (or flux) and colour, and investigate whether the results could arise from blending of clustered sources or from widespread (weak) lensing of SMGs. 
We have conducted various statistical tests and find that 
(i) red submm sources are less likely to have multiple optical associations;
(ii) the colour dependence is also observed when counting only the nearest potential counterpart to each submm source;
(iii) the colour dependence is weaker when optical associations are sought within a limited redshift range.
These observations argue against clustering (which would increase multiplicity but not affect the nearest counterpart unless it is severely blended) and in favour of lensing (which increases correlations between sources at different redshifts) as the explanation for increased offsets for red submm sources.

Furthermore, our results imply that the number of red submm sources with apparent counterparts in SDSS is significantly higher in \hatlas\ than for equivalent submm and $r$-band limits in HerMES, again probably due to a bias from lenses.
These results have important implications for the identification of optical counterparts to sources in \Herschel\ and other single-dish submm surveys, and hence for the derived redshifts and luminosity functions. The problem of mis-identifying a galaxy in a lensing structure as the counterpart to a higher redshift SMG may be more common than previously thought -- not just as a result of relatively rare strong lensing events, but also due to ubiquitous weak lensing of red SMGs.
This has an impact on the source counts in all surveys, independent of optical associations, hence it needs to be taken into account in evolutionary models for number counts.

Simulations of weak lensing by galaxies and large-scale structure suggest that in principle this effect could bias the offset distributions of SPIRE-SDSS associations,
although further work is needed to ascertain whether it is sufficiently common among 250-\mum-selected sources.
Observational work can also be used to confirm the interpretation, in particular matching statistical samples of red \Herschel\ sources to radio counterparts would help ascertain whether these are more likely to be lensed or clustered in comparison to bluer \Herschel\ sources.

\section*{Acknowledgements}
For the least-squares fitting in this work we have made use of the {\sc mpfit} package, available from \url{http://purl.com/net/mpfit} \citep{Markwardt2009}.
The authors are grateful to the anonymous referee for comments which greatly improved the clarity of arguments in the paper.
NB wishes to thank Jim Dunlop, Asantha Cooray, Shane Bussmann, Michal Micha{\l}owski, Douglas Scott and Matt Jarvis for enlightening discussions and advice, and acknowledges support from the EC FP7 SPACE project ASTRODEEP (Ref. no. 312725).
JGN acknowledges financial support from the Spanish CSIC for a JAE-DOC fellowship, co-funded by the European Social Fund, by the Spanish Ministerio de Ciencia e Innovacion, AYA2012-39475-C02-01, and Consolider-Ingenio 2010, CSD2010-00064, projects.
LD, SJM and RJI acknowledge support from the European Research Council (ERC) in the form of Advanced Investigator Program, {\sc cosmicism}.
The \Herschel-ATLAS is a project with \Herschel, which is an ESA space observatory with science instruments provided by European-led Principal Investigator consortia and with important participation from NASA. The \hatlas\ website is \url{http://www.h-atlas.org/}.
This research has made use of data from HerMES project (http://hermes.sussex.ac.uk/). HerMES is a Herschel Key Programme utilising Guaranteed Time from the SPIRE instrument team, ESAC scientists and a mission scientist.
The HerMES data was accessed through the Herschel Database in Marseille (HeDaM - http://hedam.lam.fr) operated by CeSAM and hosted by the Laboratoire d'Astrophysique de Marseille.
This work has made use of the Rainbow Cosmological Surveys Database, which is operated by the Universidad Complutense de Madrid (UCM), partnered with the University of California Observatories at Santa Cruz (UCO/Lick,UCSC).

\bibliographystyle{mn2e}
\bibliography{MyLibrary20120814}
\bsp 
\label{lastpage}
\end{document}